\documentclass[conference, letterpaper]{IEEEtran}
\IEEEoverridecommandlockouts 

\makeatletter
\def\markboth#1#2{\def\leftmark{\@IEEEcompsoconly{\sffamily}\MakeUppercase{\protect#1}}%
\def\rightmark{\@IEEEcompsoconly{\sffamily}\MakeUppercase{\protect#2}}}
\makeatother

\usepackage[mode=image|tex]{standalone} 
\usepackage[utf8]{inputenc}
\usepackage[T1]{fontenc}
\usepackage{url}
\usepackage[cmex10]{amsmath} 
\usepackage{amssymb}
\usepackage{cite}
\usepackage{graphicx}
\usepackage[caption=false, font=footnotesize]{subfig}
\usepackage{xspace} 
\usepackage{mathtools,bbm}
\usepackage{tikz,pgfplots,tkz-graph}
\usepackage{ifthen,calc}
\usepackage{listings,float}
\usepackage{epsfig,epstopdf,graphicx}
\usepackage{booktabs} 
\usepackage{hyperref}

\pgfplotsset{compat=newest}
\usetikzlibrary{positioning,matrix,shapes.multipart,shapes.misc,spy}
\newcommand{\define}{\triangleq}
\newcommand{\vect}[1]{\ensuremath{\boldsymbol{#1}}}
\newcommand{\mat}[1]{\ensuremath{\mathbf{#1}}}
\newcommand{\SNR}{\rho}
\newcommand{\EbNo}{E_\mathrm{b}/N_0}
\newcommand{\Tout}{T_\mathrm{out}}
\newcommand{\Tin}{T_\mathrm{in}}
\newcommand{\Lbit}{L_\mathrm{bit}}
\newcommand{\rdbrs}[1]{\left( #1 \right)}               
\newcommand{\rdbrsv}[2]{\left( #1 \,\middle|\, #2 \right)}              
\newcommand{\clbrsv}[2]{\left\lbrace #1 \,\middle|\, #2 \right\rbrace}  

\newcommand{\wch}{w_{\mathrm{ch}}}
\newcommand{\wmsg}{w_{\mathrm{msg}}}
\newcommand{\Hstd}{\mat{H}_{\mathrm{std}}}
\newcommand{\Hoc}{\mat{H}_{\mathrm{oc}}}
\newcommand{\Hcr}{\mat{H}_{\mathrm{cr}}}

\newcommand{\Hrr}{\mat{H}_{\mathrm{rr}}}
\newcommand{\Lmax}{L_{\mathrm{max}}}

\newcommand{\RNNSS}{RNN-SS\xspace}
\newcommand{\RNNFW}{RNN-FW\xspace}
\tikzset{%
	partial ellipse/.style args={#1:#2:#3}{%
		insert path={+ (#1:#3) arc (#1:#2:#3)}%
	}%
}%
\tikzstyle{standard BP}=[black, mark=diamond*, mark options={solid, fill=white, mark size=2.0pt}, solid]%
\tikzstyle{Nachmani}=[color=blue, mark=*, mark options={solid, fill=white, mark size=1.5pt}, dotted]%
\tikzstyle{annotation}=[fill=white]%

\newif\ifExternalBib 
\ExternalBibfalse

\begin{document}



\title{Learned Belief-Propagation Decoding with\\ Simple Scaling and
SNR Adaptation}

\author{
	\IEEEauthorblockN{
	Mengke Lian\IEEEauthorrefmark{1}, 
	Fabrizio Carpi\IEEEauthorrefmark{2}, 
	Christian H\"{a}ger\IEEEauthorrefmark{1}\IEEEauthorrefmark{3}, 
	and
	Henry D.~Pfister\IEEEauthorrefmark{1}
	\thanks{%
	The work of M.~Lian and H.~D.~Pfister was supported in part by the
	National Science Foundation (NSF) under Grant No.~1718794.  The
	work of C.~H\"ager was supported by the European Union's Horizon
	2020 research and innovation programme under the Marie
	Sk\l{}odowska-Curie grant No.~749798.  Any opinions, findings,
	conclusions, and recommendations expressed in this material are
	those of the authors and do not necessarily reflect the views of
	these sponsors. Please send correspondence to
	\texttt{henry.pfister@duke.edu}.
	}}

	\IEEEauthorblockA{\IEEEauthorrefmark{1}%
	Department of Electrical and Computer Engineering, Duke University,
	Durham, North Carolina \\
	\IEEEauthorrefmark{2}%
	Department of Engineering and Architecture, University of Parma,
	Parma, Italy \\
	\IEEEauthorrefmark{3}%
	Department of Electrical Engineering,
	Chalmers University of Technology,
	Gothenburg, Sweden }
}

\maketitle

\begin{abstract}
	We consider the weighted belief-propagation (WBP) decoder recently
	proposed by Nachmani et al.~where different weights are introduced
	for each Tanner graph edge and optimized using machine learning
	techniques. Our focus is on simple-scaling models that use the same
	weights across certain edges to reduce the storage and
	computational burden. The main contribution is to show that simple
	scaling with few parameters often achieves the same gain as the
	full parameterization. Moreover, several training improvements for
	WBP are proposed. For example, it is shown that minimizing average
	binary cross-entropy is suboptimal in general in terms of bit error
	rate (BER) and a new ``soft-BER'' loss is proposed which can lead
	to better performance. We also investigate parameter adapter
	networks (PANs) that learn the relation between the signal-to-noise
	ratio and the WBP parameters. As an example, for the $(32, 16)$
	Reed--Muller code with a highly redundant parity-check matrix,
	training a PAN with soft-BER loss gives near-maximum-likelihood
	performance assuming simple scaling with only three parameters.
\end{abstract}

\section{Introduction}

Recent progress in machine learning and off-the-shelf learning
packages have made it tractable to add many parameters to existing
communication algorithms and optimize. One example of this approach is
the weighted belief-propagation (WBP) decoder recently proposed by
Nachmani, Be'ery, and Burshtein \cite{Nachmani2016}, where different
weights (or scale factors) are introduced for each edge in the Tanner
graph. These weights are then optimized empirically using tools and
software from deep learning. Their results show that WBP can provide
significant gains over standard BP when applied to the parity-check
(PC) matrices of short BCH codes. A more comprehensive treatment of
this idea can be found in~\cite{Nachmani2018}. 

While the performance gains of WBP decoding are worth investigating,
the additional complexity of optimizing, storing, and applying one
weight per edge is significant. In this paper, we focus on
\emph{simple-scaling} models for WBP that share weights across edges
to reduce the storage and computational burden. In these models, only
 three scalar parameters are used per iteration: a message
weight, a channel weight, and a damping factor. We show that such
simple-scaling models are often sufficient to obtain gains similar to
the full parameterization. 


For WBP, the average binary cross-entropy across bit positions is used
as the optimization loss function in \cite{Nachmani2016,
Nachmani2018}. This approach is also adopted in related works, see,
e.g., \cite{Gruber2017, Bennatan2018}. On the other hand, we show that
minimizing average binary cross-entropy does not necessarily minimize
the bit error rate (BER). We propose a new loss function which can be
regarded as a ``soft'' version of BER. This loss function can lead to
performance gains when optimizing WBP with highly redundant PC
matrices, e.g., for Reed--Muller codes. 

As a last contribution, we propose a simple solution to the problem
that the optimal WBP parameters may be different for different
signal-to-noise ratios (SNRs). In particular, we use a parameter
adapter network (PAN) that learns the relation between the SNR and the
optimal WBP parameters. The usefulness of this approach is illustrated
with several examples. 

\section{Background}
\label{sec:coding}

Consider an $ (N, K) $ binary linear code $ \mathcal{C} $ defined by
an $ M \times N $ PC matrix $\mat{H}$, where $N$ is the code length,
$K$ is the code dimension, and $M \geq N-K$. We assume transmission
over the additive white Gaussian noise channel according to $y_{v} =
(-1)^{x_{v}} + z_{v}$, where $y_v$ is the $v$-th output symbol, $x_v$
is the corresponding bit in the transmitted codeword $\vect{x}$,
$z_{v}\sim \mathcal{N}(0,(2 R \EbNo)^{-1})$, and $R = K/N$ is the code
rate. We refer to $\SNR \define \EbNo$ as the signal-to-noise ratio
(SNR). 

Given any PC matrix $\mat{H}$, one can construct a bipartite Tanner
graph $\mathcal{G}=(V,C,E)$, where $ V = \{1,2,\dots, N\} \define [N]$
and $C = [M] $ are sets of variable nodes and check nodes. The edges,
$ E = \clbrsv{ (v,c) \in V \times C }{ H_{cv} \ne 0 } $, connect all
parity checks to the variables involved in them. By convention, the
boundary symbol $ \partial $ denotes the neighborhood operator defined
by $\partial v \triangleq \clbrsv{ c }{ (v,c) \in E }$ and $\partial c
\triangleq \clbrsv{ v }{ (v,c) \in E }$.

\subsection{Weighted Belief-Propagation Decoding}

WBP is an iterative algorithm that passes messages
in the form of log-likelihood ratios (LLRs) along the edges of $
\mathcal{G} $ \cite{Nachmani2016}. In the variable-to-check-node step, the pre-update
message is
\begin{align} \label{eq:LBP_vertical}
    \lambda_{v \to c}^{\prime (t)}
    = w_v^{(t)} \ell_v + \sum_{c^\prime \in \partial v \setminus c} w_{v c^\prime}^{(t)} \hat{\lambda}_{c^\prime \to v}^{(t-1)}, 
\end{align}
where $ w_{vc}^{(t)} $ is a weight assigned to the edge $(v,c)$ and $
w_v^{(t)} $ is a weight assigned to the channel message 
\begin{equation} \label{eq:LLR_def}
	\ell_v \triangleq \log \rdbrs{ \frac{ \mathrm{Pr} \rdbrsv{ y_v }{ x_v = 0 } }{ \mathrm{Pr} \rdbrsv{ y_v }{ x_v = 1 } } }
	= 4 R \rho y_v.
\end{equation}
In the check-to-variable-node step, the pre-update message is
\begin{equation} \label{eq:LBP_horizontal}
    \hat{\lambda}_{c \to v}^{\prime (t)} 
    = 2 \tanh^{-1} \rdbrs{ \prod_{v^\prime \in \partial c \setminus v} \tanh \rdbrs{ \frac{\lambda_{v^\prime \to c}^{(t)}}{2} } }.
\end{equation}
To mitigate oscillations and enhance convergence, we apply
\emph{damping} to complete the message
updates~\cite{Fossorier-istc03}. In particular, the final messages are
a convex combination of the previous value and the
pre-update value according to
\begin{IEEEeqnarray}{rCl}
    \lambda_{v \to c}^{(t)} &=& \gamma \lambda_{v \to c}^{(t-1)} +
	 (1-\gamma) \lambda_{v \to c}^{\prime (t)}, \label{eq:damping_vertical} \\
    \hat{\lambda}_{c \to v}^{(t)} &=& \gamma\hat{\lambda}_{c \to
	 v}^{(t-1)} + (1-\gamma) \hat{\lambda}_{c \to v}^{\prime (t)}, \label{eq:damping_horizontal}
\end{IEEEeqnarray}
where $ \gamma \in [0, 1] $ is the damping factor and $\lambda_{v \to
c}^{(0)} = \hat{\lambda}_{c \to v}^{(0)} = 0$ for all $(v,c) \in
E$.\footnote{Damping is referred to as ``relaxed BP'' in
\cite{Nachmani2018}, where it is studied in the context of weighted
min-sum decoding. } Finally, output LLRs are
computed as
\begin{equation} \label{eq:LBP_margin}
    m_v^{(t)} = w_v^{(t)} \ell_v + \sum_{c^\prime \in \partial c} w_{v
	 c^\prime}^{(t)} \hat{\lambda}_{c \to v^\prime}^{(t)}.
\end{equation}
The sigmoid function $ \sigma(x) = \rdbrs{ 1 + e^{-x} }^{-1} $ maps
$m_v^{(t)}$ to an estimate of the probability that $ x_v = 1 $
according to $o_v^{(t)} = 1 - \sigma ( m_v^{(t)} )$. The corresponding
hard decision is denoted by $\hat{o}_v^{(t)}$. For convenience, we
also define $o_v \define o_v^{(T)}$, $\hat{o}_v \define
\hat{o}_v^{(T)}$, and $m_v \define m_v^{(T)}$, where $T$ is the total
number of iterations.  Setting all weights to $1$ and $\gamma = 0$
recovers standard BP.

\subsection{Random Redundant Decoding}
\label{sec:rrd}


The performance of BP can be improved by using redundant PC matrices
where $M > N-K$. This can for example be realized by adding dual
codewords as rows to a standard PC matrix~\cite{Yedidia-aller02}.
Another approach, referred to as random redundant decoding (RRD), is
to use different PC matrices in each iteration~\cite{Jiang2004,
Halford-isit06}.  This can be implemented efficiently by exploiting
the code's automorphism group. In particular, let $ \mathcal{S}_N $ be
the symmetric group on $ N $ elements, i.e., $\pi \in \mathcal{S}_N$
is a permutation on $[N]$. The automorphism group of a code $
\mathcal{C} $ is defined as $\mathsf{Aut}(\mathcal{C}) \triangleq
\clbrsv{ \pi \in \mathcal{S}_N }{ \vect{x}^\pi \in \mathcal{C}, \,
\forall\, \vect{x} \in \mathcal{C} }$, where $ \vect{x}^\pi $ denotes
a permuted vector, i.e., $ x_i^\pi = x_{\pi(i)}$. RRD cascades $\Tout$
BP decoders, each run with $\Tin$ iterations, where the permuted
output LLRs of the last decoder serve as soft input for the next
decoder. A new, randomly sampled permutation $\pi_\tau \in
\mathsf{Aut}(\mathcal{C}) $ is used for each outer iteration $\tau \in
[\Tout]$. This effectively uses $\Tout$ different PC matrices while
fixing the Tanner graph for decoding.  Similar to \cite{Nachmani2018},
we consider weighted RRD by cascading several WBP decoders. 

\begin{figure}
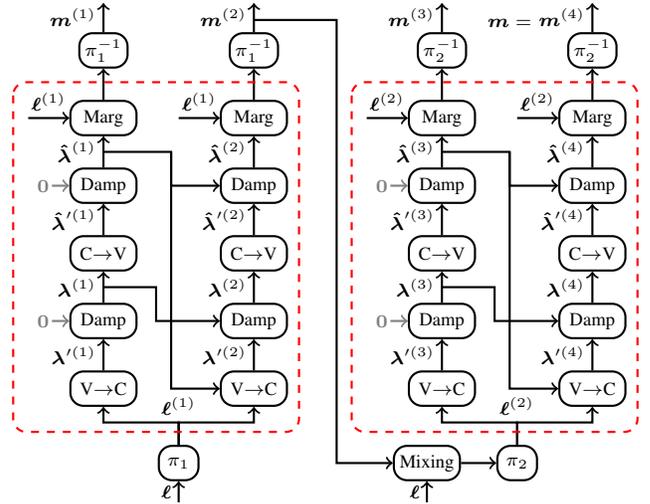

	\centering
	\includestandalone{RRD_BP_architecture}
	\vspace{-0.1cm}
	\caption{Feed-forward computation graph for RRD with $ \Tout = \Tin
	= 2 $.}
	\label{fig:RRD_BP_architecture}
\end{figure}

For RRD, the computation of output LLRs is typically modified by
introducing a scale factor before the sum in \eqref{eq:LBP_margin}
\cite[Eq.~(4)]{Jiang2004}.  We use a similar, but slightly different,
approach. In particular, the soft input to the $ \tau $-th decoder is
modified to be a convex combination of the initial channel LLRs and
the output LLRs of the $ (\tau-1) $-th decoder according to
\begin{align}
	{\vect{\ell}}^{(\tau)} = \left[ \beta \vect{\ell} + (1-\beta)
	{\vect{m}}^{((\tau-1) \Tin)} \right]^{\pi_{\tau}}, 
\end{align}
where $\beta \in [0,1]$ is referred to as the mixing factor and
$\vect{m}^{(0)} = \vect{\ell}$. Here, all BP messages and weights are
iteration-indexed consecutively for $t \in [T]$, where $T = \Tout
\Tin$. An example of the resulting feed-forward computation graph for
RRD with $\Tout = 2$ and $\Tin = 2$ is shown in
Fig.~\ref{fig:RRD_BP_architecture}. 

\section{Optimizing Weighted Belief-Propagation}

It is well known that BP performs exact marginalization when the
Tanner graph is a tree. However, good codes typically have loopy
Tanner graphs with short cycles. To improve the BP performance, one
can optimize the weights $ w_{vc}^{(t)} $ and $ w_{v}^{(t)} $ in all
iterations~\cite{Nachmani2016}. The damping and mixing factors
$\gamma$, $\beta$ can also be optimized. In the following, $\vect{o}
=f(\vect{y}; \theta)$ denotes the entire WBP mapping, where $\theta$
comprises all parameters. 

\subsection{Deep Learning via Stochastic Gradient Descent}

In \cite{Nachmani2016}, the authors propose to optimize $\theta$ using
stochastic gradient descent (SGD) (or a variant thereof) based on many
codeword--observation pairs $(\vect{x},\vect{y})$. In particular, the
empirical loss $\mathcal{L}_A (\theta)$ for a finite set $A \subset
\mathcal{C} \times \mathbb{R}^N$ of pairs is defined by
\begin{align}
\mathcal{L}_A (\theta) \triangleq \frac{1}{|A|}
\sum_{(\vect{x},\vect{y})\in A} L \big( \vect{x}, f(\vect{y};\theta) \big),
\end{align}
where $L(\vect{a}, \hat{\vect{a}})$ is the loss associated with
returning the output $\hat{\vect{a}}$ when $\vect{a}$ is correct.
Mini-batch SGD then uses the parameter update $\theta_{i+1} = \theta_i
- \alpha \nabla_{\theta} \mathcal{L}_{B_i} (\theta_i)$, where $\alpha$
is the learning rate and $B_i$ is the mini-batch used in the
$i$-th step. Due to channel and decoder symmetry, transmission of the
all-zero codeword $\vect{x} = \vect{0}$ can be assumed for the
optimization \cite{Nachmani2016}. 

\subsection{Optimization Loss Function}

For supervised classification problems, one typically uses
cross-entropy loss. However, since the number of classes (i.e.,
codewords) scales exponentially with the block length, it is more
practical to assume that the overall loss is the average of
\emph{bit-wise} losses according to
\begin{align}
	\label{eq:loss}
	L(\vect{x}, \vect{o}) = \frac{1}{N} \sum_{v=1}^N \Lbit(x_v, o_v), 
\end{align}
where $\Lbit$ is a bit-wise loss function. For optimizing WBP, binary
cross-entropy $\Lbit(a,b) = -\log(b^a (1-b)^{1-a})$ is used in
\cite{Nachmani2016, Nachmani2018}. However, our experiments show that
minimizing \eqref{eq:loss} using binary cross entropy does not
necessarily minimize the BER. To see why this may be the case, note
that the negative bit success rate (per codeword) can be written as
\begin{align}
	\label{eq:bit_success}
	\frac{1}{N} \sum_{v=1}^N \Lbit(x_v,
	\hat{o}_v) = - \frac{1}{N} \sum_{v=1}^N \hat{o}_v^{x_v} (1-\hat{o}_v)^{1-x_v}, 
\end{align}
where $\Lbit(a,b) = -b^a (1-b)^{1-a}$. On the other hand, inserting
binary cross entropy into \eqref{eq:loss} leads to
\begin{align}
	\label{eq:cross_entropy}
	L(\vect{x}, \vect{o}) = - \log \left( \prod_{v=1}^{N}
	 {o}_v^{x_v} (1-{o}_v)^{1-x_v}
	 \right)^{\frac{1}{N}}. 
\end{align}
Besides the log, the main difference between \eqref{eq:bit_success}
and \eqref{eq:cross_entropy} is that arithmetic mean is used instead
of geometric mean.

We propose a new loss function, where $\Lbit(a,b) = (1-b)^a b^{1-a}$ is
used in \eqref{eq:loss}. This can be regarded as a ``soft'' version of
BER since $(1-b)^a b^{1-a}$ for binary variables corresponds to $a$ XOR
$b$, i.e., $\Lbit(a,b)$ indicates a bit error.  We refer to the
resulting loss function as soft-BER. Tab.~\ref{tab:loss} summarizes
the different binary loss functions and their simplification for the
all-zero codeword. 




\newcommand{\tablehighlight}[1]{{#1}}
\begin{table}
	\caption{Comparison of bit-wise loss functions.}
	\label{tab:loss}
	\centering
	\renewcommand{\arraystretch}{1.2}
	\begin{tabular}{ccc}
		\toprule
		\tablehighlight{name} & $\Lbit(a,b)$ & $\Lbit(0,b)$  \\ 
		\midrule
		binary cross-entropy & $-\log( b^a (1-b)^{1-a} )$ & $-\log(1-b)$ \\
		negative soft bit success & $- b^a (1-b)^{1-a}$ & $-(1-b)$ \\
		soft bit error            & $ (1-b)^a b^{1-a}$ & $b$ \\
		\bottomrule
	\end{tabular}
\end{table}

\subsection{Multi-Loss Optimization}
\label{sec:multi-loss}

The optimization behavior for WBP can be improved by using a
multi-loss function \cite{Nachmani2016, Nachmani2018} (see also
\cite{Bennatan2018})
\begin{align} 
	\label{eq:multi-loss}
	L( \vect{x},  \{\vect{o}^{(t)}\}_{t=1}^T) \triangleq \frac{1}{
	\sum_{t=1}^{T} \eta^{T-t} }
	\sum_{t=1}^{T} \eta^{T-t} L (\vect{x}, \vect{o}^{(t)}), 
\end{align}
where $\eta \in [0, 1]$ is a discount factor. Multi-loss optimization
takes into account the output after every iteration which helps to
increase the magnitude of gradients corresponding to earlier
iterations. We found that, rather than using a fixed discount factor
as in \cite{Nachmani2016, Nachmani2018, Bennatan2018}, it is
beneficial to decay $\eta$ during SGD, i.e., gradually moving from
$\eta = 1$ (where the outputs of all BP iterations are considered with
equal importance) towards $\eta = 0$ (where only the last BP iteration
is taken into account). 

\subsection{Weight Sharing}

Excluding the damping/mixing factors, the total number of weights is
$T(|E|+N)$ and we refer to this case as the fully-weighted (FW)
decoder. In order to reduce the optimization complexity, one can share
the weights, e.g., as follows:

\begin{itemize}
	\item Temporal weight sharing (across decoding iterations), i.e.,
		\begin{equation*}
			w_{vc}^{(t)} \equiv w_{vc}, \qquad w_v^{(t)} \equiv w_v, 
			\qquad \forall\, t \in [T],
		\end{equation*}
		is referred to as \RNNFW, due to the similarity with
		recurrent neural networks (RNNs) \cite{Nachmani2016}. 

	\item Spatial weight sharing (across edges), i.e.,
		\begin{equation*}
			w_{vc}^{(t)} \equiv \wmsg^{(t)}, \qquad w_v^{(t)} \equiv \wch^{(t)}, \qquad \forall\, (v, c) \in E,
		\end{equation*}
		gives the simple-scaling (SS) model with two weights per iteration:
		one message and one channel weight. 

	\item Temporal \emph{and} spatial weight sharing gives two
		parameters in total. This is referred to as \RNNSS. 

\end{itemize}

It is shown in \cite{Nachmani2016, Nachmani2018} that the \RNNFW
structure gives similar gains as the FW decoder, i.e., there is little
improvement when making parameters iteration-dependent. We further
show that the \RNNSS structure incurs little to no performance penalty
in many cases. 

\subsection{Training SNR and Parameter Adapter Network}

In general, the optimal WBP parameters may be different for different
SNRs \cite{Nachmani2018}. On the other hand, optimizing WBP separately
for each SNR and storing the resulting weights is impractical if the
set of possible SNRs is large or infinite. One general approach is to
instead optimize assuming a range of different training SNRs
\cite{Nachmani2016, Nachmani2018}. This leads to parameters that
achieve a compromise between different channel conditions.

\begin{figure}[t]
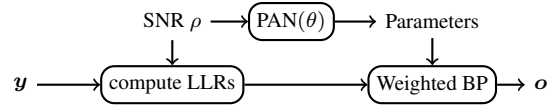

	\centering
	\includestandalone{PAN}
	\vspace{-0.1cm}
	\caption{Block diagram illustrating the parameter adapter network
	(PAN).}
	\label{fig:adaptive_learned_BP}
\end{figure}

We propose a different approach where a PAN is used to learn the
relation between the SNR $\rho$ and the corresponding optimal
parameters.\footnote{We assume perfect knowledge of the SNR. This
knowledge is also required implicitly to compute channel LLRs. In
practice, SNR is typically estimated and the SNR estimate can then be
used as the input to the PAN.} Once trained, the PAN can be used to
adaptively choose the best parameters for WBP corresponding to the
channel conditions. The basic idea is illustrated in
Fig.~\ref{fig:adaptive_learned_BP}. In general, one can choose any
structure to construct the PAN, e.g., a vanilla neural network. It is
also possible to make only a subset of parameters SNR-adaptive. 

In this paper, we use several shallow neural networks with one hidden
layer of dimension $20$ and output dimension $1$ to model the
SNR-dependency for each WBP parameter separately. ReLU activations are
used for the hidden layer. The output layer uses sigmoid activations
to ensure that the parameters satisfy their domain constraints, e.g.,
the damping factor is in the range $[0,1]$. For regular weights, we
further scale the sigmoid outputs by $10$ to increase the range to
$[0, 10]$. As an example, for WBP with the \RNNSS structure including
damping, there are three parameters $w_{\text{msg}}$, $w_{\text{ch}}$,
and $\gamma$. Thus, the PAN describes an SNR-parameterization
according to $\text{PAN}(\SNR) = [ w_{\text{msg}}(\SNR), \,
w_{\text{ch}}(\SNR), \, \gamma(\SNR) ] \in [0,10]^2 \times [0,1]$.

\section{Numerical Results}



The various decoding architectures in this paper are implemented in
the PyTorch framework and optimized using the RMSprop optimizer which
is a variant of mini-batch SGD. Each mini-batch contains $100$
observation pairs and the SNR for each pair is chosen from $10$
equidistant points in the interval $[1\,\text{dB}, 8\,\text{dB}]$ such
that exactly $10$ pairs have the same SNR. The discount decay for the
multi-loss optimization is implemented by starting with an initial
discount factor $\eta = 1$ and multiplying $\eta$ by $0.5$ after every
$5000$th SGD step. The same schedule is used to decay the learning
rate, starting from $\alpha = 10^{-3}$ and using a decay rate of $0.8$
instead of $0.5$. To avoid numerical issues, a gradient clipping
threshold of $ 0.1 $ is applied and the absolute values of the LLRs
${\lambda}_{v \to c}^{(t)} $ are clipped into the range
$[-\log(\tanh(\Lmax/2)), \Lmax]$ with $\Lmax = 15$. 

The following Reed--Muller (RM) and Bose--Chaudhuri--Hocquenghem (BCH)
codes are considered: 

\begin{itemize}
	\item RM$(32,16)$ with standard PC matrix $\Hstd$ (size $16 \times
		32$) and overcomplete PC matrix $\Hoc$ ($620 \times 32$) whose
		rows are all minimum-weight dual codewords, see \cite{Santi2018}

	\item BCH$(63, 36)$ with cycle-reduced PC matrix $\Hcr$ ($27 \times
		63$) and right-regular PC matrix $\Hrr$ ($27 \times 63$), see
		\cite{Helmling2017}


	\item BCH$(127, 64)$ with cycle-reduced PC matrix $\Hcr$ ($63
		\times 127$), see \cite{Helmling2017} 

\end{itemize}

Ordered statistics decoding (OSD) is used as a benchmark whose
performance is close to maximum-likelihood \cite{Fossorier1995}. 

\subsection{Comparison of Loss Functions}

\begin{figure}[t]
	\centering
	\vspace{-0.40cm}
	\subfloat[RM$(32,16)$, overcomplete $\Hoc$]{%
		\includestandalone{loss_RM}%
	}%
	\subfloat[BCH$(63,36)$, cycle-reduced $\Hcr$]{%
		\includestandalone{loss_BCH}%
	}%
	\vspace{0.05cm}
	\caption{Comparison of loss functions for \RNNSS
	with $ w_{\mathrm{ch}} = 1 $, $\gamma = 0$, and 
	$T = 3$. The SNR is $\EbNo = 3\,$dB in (a) and $\EbNo =
	7\,$dB in (b). }
	\label{fig:loss_comparison}
\end{figure}

We start by considering two \RNNSS structures with fixed $\wch = 1$
and $\gamma = 0$ (i.e., no damping): (a) RM$(32,16)$ with $\Hoc$ and
(b) BCH$(63,36)$ with $\Hcr$. The different loss functions for $T=3$
are plotted in Fig.~\ref{fig:loss_comparison} as a function of
$\wmsg$, which is the only trainable parameter. For the RM code,
cross-entropy has a sharp minimum at $\wmsg \approx 0.05$, whereas
soft-BER overlaps with BER and has a flat minimum at $\wmsg \approx
0.15$. For the BCH code, the minima for cross-entropy, soft-BER, and
BER all occur close to each other, but at slightly different
locations.

In order to explain the distinct behavior of cross-entropy in
Fig.~\ref{fig:loss_comparison}(a), note that if a bit is decoded
incorrectly, binary cross-entropy gives a penalty close to the
magnitude of the output LLR $|m_v|$. This is problematic in cases
where the decoder is wrong, but very sure about its decision. Indeed,
this behavior is characteristic for BP with highly redundant PC
matrices and such failure cases tend to dominate the average loss.
This effect is even more pronounced for large $T$ since the average
LLR magnitude tends to grow with the iteration number. 

The results in Fig.~\ref{fig:loss_comparison} show that, in general,
neither cross-entropy nor soft-BER are guaranteed to minimize BER. All
scenarios in this paper were optimized using both functions. We found
that they give comparable results, with the exception of highly
redundant PC matrices where soft-BER is preferable. 


\subsection{Reed--Muller Codes}

\begin{figure}[t]%
	\centering%
	\includestandalone{RM_32_16}%
	\vspace{-0.20cm}%
	\caption{Results for RM$(32, 16)$ with $\Hoc$ ($T = 5$) and $\Hstd$
	($T=20$).}%
	\label{fig:reed_muller}%
\end{figure}%

Results for RM$(32, 16)$ assuming both \RNNFW and \RNNSS structures
are shown in Fig.~\ref{fig:reed_muller}. For $\Hstd$ with $T = 20$
iterations, simple scaling results in a performance loss of up
to $0.3\,$dB. Damping gives considerable performance improvements in
both cases, at the expense of additional computational complexity and
storage requirements. For $\Hoc$ with $T=5$, the \RNNSS structure is
sufficient to achieve close-to-optimal performance and the overlapping
results for \RNNFW are omitted. For this case, we note that the
optimization with soft-BER gives lower BER than for CE, as
expected from the discussion in the previous subsection. 

\subsection{BCH Codes}

\begin{figure*}[t]
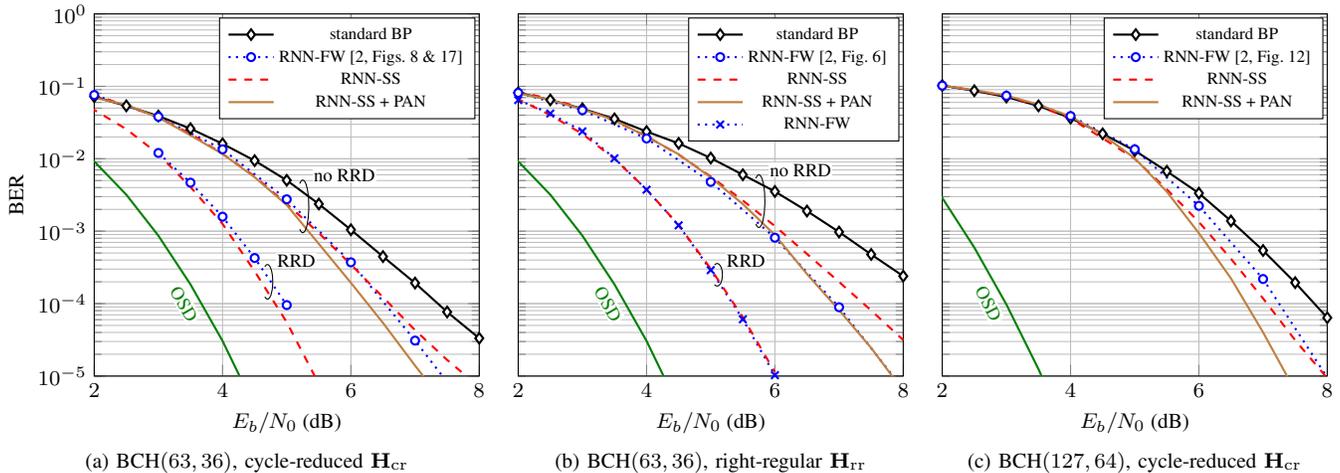

	\vspace{-0.40cm}
	\centering%
	\subfloat[BCH$(63, 36)$, cycle-reduced $\Hcr$]{\includestandalone{BCH_63_36_cr}}%
	$\,$
	\subfloat[BCH$(63, 36)$, right-regular $\Hrr$]{\includestandalone{BCH_63_36_rr}}%
	$\,$
	\subfloat[BCH$(127, 64)$, cycle-reduced $\Hcr$]{\includestandalone{BCH_127_64_cr}}%
	\vspace{0.05cm}
	\caption{Results for BCH codes assuming $\gamma=0$ (i.e., no
	damping), $T=5$ iterations (no RRD) and $\Tin=2$, $\Tout=30$ for RRD. 
	Data points are
	extracted directly from the relevant figures in
	\cite{Nachmani2018}. Note that our results for
	BCH$(127,64)$ are not directly comparable to \cite{Nachmani2018} because of
	potentially different PC matrices. }
	\label{fig:BCH}
	\vspace{-0.2cm}
\end{figure*}

For the BCH codes, the parameters are chosen to facilitate a direct
comparison with \cite{Nachmani2018}. In particular, we fix $T=5$ and
$\gamma = 0$, i.e., no damping is used. Results for BCH$(63, 36)$ with
$\Hcr$ are shown in Fig.~\ref{fig:BCH}(a), where we compare with the
best results in \cite[Fig.~8]{Nachmani2018} for the same parameters.
The \RNNSS with two trainable parameters achieves similar gains as the
\RNNFW in \cite{Nachmani2018} for BERs $> 10^{-4}$.  For lower BERs,
the performance starts to deviate. The situation can be improved by
making the parameters SNR-adaptive using the proposed PAN approach. In
this case, the performance improves markedly for high SNRs. This is
due to the fact that training over a range of SNRs without the PAN
tends to focus almost exclusively on low-SNR/high-BER regions. Similar
observations can be made for the same code with $\Hrr$, as shown in
Fig.~\ref{fig:BCH}(b). In this case, \RNNFW performs slightly better
than \RNNSS for some SNRs, i.e., two parameters are not sufficient to
obtain the full gain. Finally, results for BCH$(127, 64)$ with $\Hcr$
are shown in Fig.~\ref{fig:BCH}(c). We caution the reader that these
results are not directly comparable because our standard BP performs
better than what is shown in \cite[Fig.~12]{Nachmani2018}. This is
likely due to different cycle-reduced PC matrices. However, we were
not able to improve upon the shown \RNNSS results using \RNNFW, which
indicates that the simple-scaling approach is also sufficient in this
case. 


We also consider weighted RRD for BCH$(63, 36)$ where $\Tin = 2$,
$\Tout = 30$, and the mixing factor is treated as an additional
optimization parameter. Results for $\Hcr$ are shown in
Fig.~\ref{fig:BCH}(a) and we compare to the corresponding results in
\cite[Fig.~12]{Nachmani2018} labeled as ``mRRD-RNN(1)''. We obtain
slightly better performance using RNN-SS even without a PAN. This can
be attributed to the improved training methodology, particularly the
discount decay for the multi-loss optimization. For $\Hrr$, no RRD
results are available in \cite{Nachmani2018} and we compare to our own
results. Both RNN-FW and RNN-SS give virtually the same performance as
shown in Fig.~\ref{fig:BCH}(b). Additional simulation results for
larger $T$ can be found in \cite{Lian2018itw}.

\section{Discussion and Conclusion}
\label{sec:conclusion}

In this paper, we have considered WBP decoding of short Reed--Muller
and BCH codes. Our experiments support the observations
in~\cite{Nachmani2016,Nachmani2018} that optimizing WBP can provide
meaningful gains. In addition, we have shown that simple-scaling
models with fewer parameters are often sufficient to achieve gains
similar to the full parameterization. This can lead to a considerably
simpler optimization procedure and greatly reduce complexity, e.g., in
terms of storage requirements. In general, the performance loss
incurred by simple scaling depends on the employed PC matrix. Small
penalties were observed for matrices with highly irregular degree
distributions (e.g., $\Hstd$ for RM(32,16) or $\Hrr$ for BCH(63, 36)),
whereas the loss appears to be negligible if the degree distribution
is regular ($\Hoc$ for RM(32,16)) or RRD is employed.

It was also shown that choosing a suitable loss function for the
optimization is scenario-dependent. For highly redundant PC matrices,
it was found that binary cross-entropy penalizes too hard on bit
errors where the decoder is very sure about its decision. In such
cases, optimizing with the proposed soft-BER loss leads to better
performance. Lastly, we built on the observation in
\cite{Nachmani2018} that the optimal WBP parameters are SNR-dependent
and proposed a simple solution based on parameter adapter networks.
This approach allows us to learn optimal parameters for multiple SNRs
in a single training process without trading off performance between
channel conditions. 

\vspace{-0.1cm}
\ifExternalBib



\else

\fi

\end{document}